\documentclass[12pt,preprint]{aastex}

\slugcomment{To appear in the Publications of the Astronomical Society of the Pacific}

\shorttitle{NASA Exoplanet Archive}
\shortauthors{Akeson et al.}

\begin{document}


\title{The NASA Exoplanet Archive: Data and Tools for Exoplanet Research}


\author{R.L. Akeson\altaffilmark{1}, X. Chen\altaffilmark{1}, D. Ciardi\altaffilmark{1}, M. Crane\altaffilmark{1}, J. Good\altaffilmark{1}, M. Harbut\altaffilmark{2}, E. Jackson\altaffilmark{2}, S.R. Kane\altaffilmark{1}, A.C. Laity\altaffilmark{1}, S. Leifer\altaffilmark{1}, M. Lynn\altaffilmark{2}, D.L. McElroy\altaffilmark{1}, M. Papin\altaffilmark{1}, P. Plavchan\altaffilmark{1}, S.V. Ram\'{i}rez\altaffilmark{1}, R. Rey\altaffilmark{2}, K. von Braun\altaffilmark{1}, M. Wittman\altaffilmark{2}, 
M. Abajian\altaffilmark{1}, B. Ali\altaffilmark{2}, C. Beichman\altaffilmark{1}, A. Beekley\altaffilmark{1}, G.B. Berriman\altaffilmark{1}, S. Berukoff\altaffilmark{1}, G. Bryden\altaffilmark{3}, B. Chan\altaffilmark{1}, S. Groom\altaffilmark{2}, C. Lau\altaffilmark{1}, A.N. Payne\altaffilmark{1}, M. Regelson\altaffilmark{1}, M. Saucedo\altaffilmark{1}, M. Schmitz\altaffilmark{2}, J. Stauffer\altaffilmark{2},
P. Wyatt\altaffilmark{1}, A. Zhang\altaffilmark{2} }

\altaffiltext{1}{NASA Exoplanet Science Institute, California Institute of Technology,
Pasadena, CA, 91125}
\altaffiltext{2}{Infrared Processing and Analysis Center, California Institute of
Technology, Pasadena, CA, 91125}
\altaffiltext{3}{Jet Propulsion Laboratory, California Institute of
Technology, Pasadena, CA, 91108}

\begin{abstract}
We describe the contents and functionality of the NASA Exoplanet Archive, a database and tool set funded by NASA to support astronomers in the exoplanet community. The current content of the database includes interactive tables containing properties of all published exoplanets, Kepler planet candidates, threshold-crossing events, data validation reports and target stellar parameters, light curves from the Kepler and CoRoT missions and from several ground-based surveys, and spectra and radial velocity measurements from the literature.  Tools provided to work with these data include a transit ephemeris predictor, both for single planets and for observing locations, light curve viewing and normalization utilities, and a periodogram and phased light curve service. The archive can be accessed at http://exoplanetarchive.ipac.caltech.edu.
\end{abstract}


\section{Overview and Archive Goals}

Since the announcement of the planet around 51 Peg \citep{may95}, the exoplanet field of astronomy and the rate of exoplanet discovery continues to increase. The number of confirmed and candidate exoplanets, the methods used to discover and characterize these exoplanets, and the volume of observations from space missions and ground-based telescopes have resulted in a diverse range of reported planetary properties. The histogram shown on the right in Figure 1 shows the increasing rate of discovery, as well as how the different techniques contribute to the discoveries, with radial velocity and transit techniques clearly being the dominant contributors. Many members of this exciting field have worked to collect information on exoplanets and make this information readily available to both astronomers working in this field and to interested members of the public.  For example, the Extrasolar Planet Encyclopedia \citep[][\url{http://exoplanet.eu/}] {sch11} has been operating since 1995 and includes a comprehensive list of confirmed and retracted exoplanets, and the Exoplanet Orbit Database \citep[][\url{http://exoplanets.org}]{wri11} maintains a database of planets with well determined orbital parameters. Other resources for the exoplanet community include archives from the dedicated space missions, the Kepler archive at MAST (Mikulski Archive for Space Telescopes; \url{http://archive.stsci.edu/kepler/}) and the CoRoT archive at the IAS (\url{http://idoc-corot.ias.u-psud.fr/}), and services such as the Exoplanet Transit Database \citep[][\url{http://var2.astro.cz/ETD/}]{pod10}.

\begin{figure}[h!]
\includegraphics[width=6.5in]{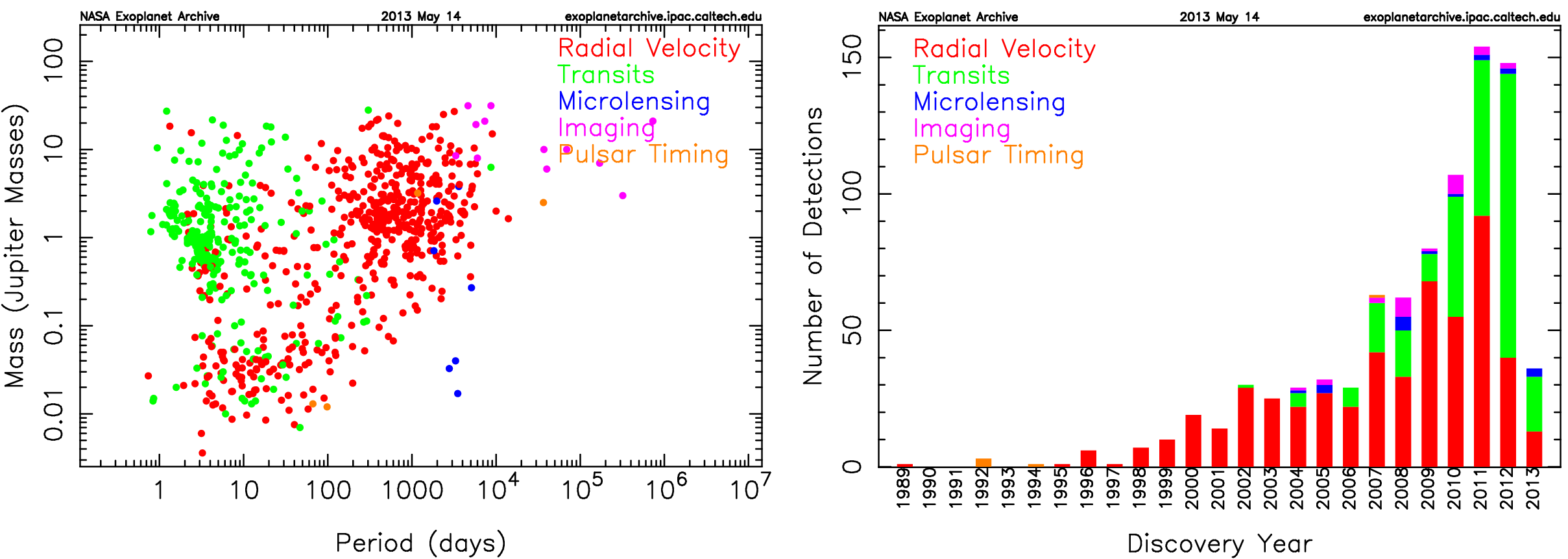}
\caption{\footnotesize Examples of pre-generated plots from the Exoplanet Archive. Right: Histogram of the exoplanets discovered as a function of time. The different techniques used are indicated by different colors. Left: Plot of planetary mass vs. orbital period for the confirmed exoplanets. The plotted points are color-coded by the method of detection in each case.  Additional plots are available online.
\label{fig:pregen}
}
\end{figure}

In this paper, we describe the NASA Exoplanet Archive (referred to as the "Exoplanet Archive"), an online astronomical exoplanet and stellar catalog and data service provided to the astronomical community to assist in the search for and characterization of exoplanets and their host stars.  The Exoplanet Archive is funded by NASA and developed and operated by the NASA Exoplanet Science Institute (NExScI) at Caltech. This archive combines a database of confirmed exoplanet and host star properties with key public data sets from space and ground-based surveys and provides quantitative analysis tools to work with these data.  Examples of the data included are stellar parameters (positions, magnitudes, and temperatures), exoplanet parameters (masses and orbital parameters), and discovery/characterization data (published radial velocity curves, photometric light curves, images, and spectra). The contents of each data set are fully described as part of the archive documentation.
The Exoplanet Archive also includes over 2.9 million light curves, including public data from the Kepler \citep{bor10} and COnvection ROtation and planetary Transits \citep[][CoRoT]{auv09} space missions and several ground-based surveys, and products from the Kepler data pipeline. This archive includes both data that are available elsewhere (e.g. the confirmed planet list and the Kepler light curves) and data that are released through the archive (e.g. Kepler data validation reports and Kepler Object of Interest lists; see Section 2.2). Additionally, the archive provides extensive documentation on its data and tools.

Our goal is to facilitate exoplanet research by providing a single location and a consistent set of tools to work with these data. This unique combination of data and tools allow users to:

\begin{itemize}
\item Compare stellar and planetary physical and orbital values published by different detection methods;
\item Develop target lists for new observations based on previously published results;
\item Develop different algorithms for transit detection or variability classification using complete light curve data sets; for instance, to enable the detection of planets not reported in the original study;
\item Extend the time baseline for transit studies by combining data sets containing the same stars, leading to increased detection efficiency and enhanced potential to conduct transit timing studies;
\item Enable additional science not pursued in the original survey, such as studies of eclipsing binary and other variable stars or time-dependent phenomena, stellar atmospheres (rotation, flares, spots, etc.), asteroseismology and intrinsic stellar variability, as well as serendipitous discoveries such as photometric behaviors of supernovae progenitors.
\end{itemize}

Much of the content and infrastructure for the Exoplanet Archive was adapted from the NASA Stellar and Exoplanet Database\citep[NStED;][]{von09,ram09}.  The stellar and planetary data, Kepler candidate data and pipeline results, and all light curve metadata are stored in a relational database, while the light curves, Kepler data validation reports and some additional data are stored in a file system.  Users interact with these data either through web-based tools or by using wget or HTML calls to query and return data directly. This paper describes the content and functionality of the Exoplanet Archive at the time of writing.  As the archive is still undergoing substantial updates to both the data content and the tools, we encourage interested readers to visit the website at \url{http://exoplanetarchive.ipac.caltech.edu}for the most up-to-date information.

\section{Data Content}

The main source of exoplanet and host star data in the Exoplanet Archive is the refereed literature. These data are vetted by a team of astronomers and are linked back to the original literature reference. Data are searchable either for an individual star or by stellar and planetary properties. The Exoplanet Archive offers direct access to frequently accessed data sets via interactive tables, which allow data to be sorted and filtered. These data sets include a list of all known planets and hosts and a list of all Kepler planet candidates, confirmed planets and false positives. The database and interface design allow for the storage and display of multiple values for parameters; this is currently available for the stellar properties and will be expanded to include planetary properties. Asymmetric uncertainties and limits are stored and displayed where appropriate.

\subsection{Exoplanets}

One Exoplanet Archive objective is to compile a database of exoplanet values for both new exoplanets and updated parameters.  We do this by monitoring submissions via the journal pages and the LANL astro-ph server\footnote{\url{http://xxx.lanl.gov/archive/astro-ph}} and extracting exoplanet parameter information directly from accepted papers in the refereed literature. The updates are performed on a weekly basis and involve internal validation of the data against the literature values.  New exoplanet information is generally available on the website within 1 to 2 weeks.  

The interactive table for confirmed planets contains over 75 planetary and stellar physical and orbital parameters, along with their uncertainties and limits. Additionally, an overview page is available for each confirmed planetary system, which contains not only the stellar and planetary values available in the interactive table, but also any additional published values the archive has collected.  For instance, many of the brightest stellar hosts have multiple values for the same stellar property.  Any additional files such as spectra, images or radial velocity files (see \S \ref{add:data}) are also available from this overview page.  The overview page can be accessed either from the Search for a Planet or Stellar System link on the home page or from the links in the Confirmed Planets interactive table (\S \ref{interactive}).   Updates may be made to the listed parameters if newly published values include additional parameters or are more precise.

The exoplanet discoveries over the last two decades have revealed a continuum of planetary masses that stretches from the planetary realm into objects with the mass of brown dwarfs. Since there are no universally adopted criteria for establishing a dividing line between planetary and other sub-stellar objects, the Exoplanet Archive adopts the following criteria for inclusion of an exoplanet: (1) have a mass (or minimum mass) estimate that is equal to or less than 30 Jupiter masses, (2) the properties of the planet are described in the peer-reviewed literature, and (3) sufficient follow-up observations and validation have been undertaken to deem the possibility of the object being a false positive as unlikely. Decisions on inclusion of exoplanets considered to be tentative or controversial and on which parameter values are cited are made by the NExScI scientists at the Exoplanet Archive and occasionally change as new information is available on a particular planet or technique.
In the case of multiple sets of values available in the literature for a given planet or host star, this decision process includes which reference to use in the
confirmed planets table and is based on the uncertainities and completeness of the published data sets.    A link to the selected reference
is available via the overview page for each planet and we note that differences between values in our confirmed planets table and in the
other exoplanet resources in \S 1 are due in part to selection of different published values.
Exoplanets may also be removed from the confirmed table if new observations or analysis demonstrate that the original claim is no longer supported.  An example is the case of VB 10 b which was first reported using astrometry \citep{pra09}, but later radial velocity data \citep{bea10} did not support the claimed planet properties. Users who wish to apply more stringent criteria on the mass limit can easily do so using the filtering mechanism in the interactive table.

Exoplanet host stars usually have several names and aliases from a variety of catalogs and the Exoplanet Archive records and displays these names on the overview page. However, the name that is displayed by default (i.e. in the Confirmed Planets interactive table) depends on the name most commonly used in the literature or the name used in the discovery paper. For example, bright exoplanet host stars are most commonly referred to by their entry in the Henry Draper (HD) catalog \citep{can18}. Fainter stars are usually referred to by their designation assigned by the survey that discovered the planet. This is typical of surveys that discover planets with the transit method, such as Super Wide Angle Search for Planets (SuperWASP), the Hungarian Automated Telescope Network (HATNet), and Kepler. 
A list of aliases is available on the overveiw page.
If the exoplanet is known to orbit a single member of a multiple stellar system (e.g. alpha Cen B or GJ 676A), the stellar host name includes the stellar component letter. For the planet designation, we follow the convention of lower-case letters.

\subsection{Kepler Pipeline Data}

The Kepler mission is surveying over 150,000 stars in a search for Earth-sized planets via the transit method \citep{bor10}. The Kepler pipeline examines the light curves for all objects to identify possible transit events, and then performs complex multi-quarter and multi-event modeling \citep{jen10}. Each transit-like event detected by the pipeline with a signal-to-noise ratio greater than 7 constitutes a threshold-crossing event (TCE).  These TCEs  are further studied and characterized to identify planet candidates, eclipsing binaries, and false positives.  The remaining objects are placed on the Kepler Object of Interest (KOI) list, and are subjected to follow-up observations and further analysis to confirm or validate their planetary status. The Kepler mission utilizes two archives to provide data to the community.  The light curves, pixel data files, cotrending basis vectors and other engineering data are available at the Mikulski Archive for Space Telescopes (MAST; \url{http://archive.stsci.edu/kepler/}), while the products of the pipeline, including the TCE list, the KOI lists (which contain both planet candidates and false positives), and the data validation reports are available in the Exoplanet Archive.  MAST and the Exoplanet Archive coordinate directly to ensure all Kepler data products will be archived permanently for the community and that users have access to the most current information needed to analyze and interpret Kepler data.  For instance, the Exoplanet Archive downloads the Kepler light curves from MAST so users can see the light curves associated with a given KOI or TCE. Conversely, the MAST archive maintains a list of current KOIs by querying the Exoplanet Archive once a day.

The Exoplanet Archive presents data related to KOIs in an integrated and interactive table. This table includes stellar parameters (effective temperature, gravity, etc.) and transit parameters (periods, depths, durations, etc.) and derived planet properties (radius, etc.). Parameters from the Kepler Input Catalog (KIC) have been added, as well as designating KOIs that have been confirmed as planets. During the Kepler extended mission, multiple version of the KOI lists will remain available.  New KOI lists will be made available in the archive during the process of vetting the KOIs and dispositioning them into categories of planet candidate and false positive.   Once all work on a given table is completed, this table is marked as Done in the vetting status, indicating that no more changes will be made.  Users interested primarily in completeness work should use tables marked as Done. Users interested in the most recent values for a given KOI should use the most recent quarter list, but should be aware that values and dispositions will change as the Kepler team continues to work on this list. As of May 2013, the KOI lists are from Kepler Quarters 1-6 \citep[][Bryson et al, in preparation]{bat13}, Quarters 1-8 (Burke et al, in preparation), Quarters 1-12 and a cumulative list. As the pipeline detection of individual KOIs may change as different data are processed, the cumulative KOI table compiles historical information from the individual KOI tables to provide the most accurate dispositions and stellar and planetary information in one place and includes all the KOIs from the quarter-based lists. The KOI lists are available as tabs within a single interactive table, so the user can access each list of KOIs with self-consistent parameters and monitor the evolution of parameters of an individual KOI as more quarters of data are added into the transit modeling analysis.   The pipeline overview page for each KOI includes the parameter values from each list containing that KOI.

The Kepler pipeline identifies tens of thousands of TCEs each time the pipeline is run.  The pipeline output includes transit modeling parameters, centroid results, and several statistical values for each TCE, in addition to summary and detailed reports (in PDF format), all of which are available to users via the Exoplanet Archive.  These data products are available as an interactive table for the TCEs and on the pipeline overview page for each KOI.  The Exoplanet Archive serves additional information related to the KOIs, including results of variability studies \citep{deb11} and eclipsing binary catalogs \citep{prs11}.  This additional information, as well as all pipeline values, is included on the pipeline overview page that is generated for each Kepler candidate.  It can be accessed from the Kepler candidate's entry in the interactive table.  The data validation reports are available from the TCE interactive table, the overview page or via the wget scripts (\S \ref{bulkdl}).  The TCE data currently available correspond to the pipeline run using the quarter 1 through 12 (Q1-12) data as input. Additionally, the archive includes stellar parameters for all targets observed by Kepler for the purpose of finding planets (over 190,000 stars).  Currently the available stellar data are for the Q1-12 pipeline run matching the TCE table and are available in an interactive table and via the API.

\subsection{Kepler and CoRoT Light Curves}

The primary mission archive for Kepler light curves and pixel data is provided by MAST.  The Exoplanet Archive provides access to the public Kepler light curves, both short and long cadence, as they are released to the community.  The Kepler light curves for all $\sim$190,000 objects in the Kepler Target Catalog are accessible via a custom search page, and light curves for KOIs and confirmed Kepler planets can also be accessed via the Kepler interactive table.  Tools are provided to visualize, manipulate, normalize and stitch time series as discussed in \S \ref{timeseries}.  Releases of new quarters are made approximately on a three-month schedule. Searches can be made by astrophysical or observational parameters tabulated by the Kepler project, or by the KIC ID. Time series from multiple quarters are tied to each KIC ID and can be downloaded together for offline analysis. As a note of caution, the Exoplanet Archive does not adjust the time system of the Kepler time series as provided by the Kepler project pipeline, and there can be timing offset errors\footnote{\url{http://archive.stsci.edu/kepler/timing\_error.html}}.

The Exoplanet Archive is also the U.S. portal for CoRoT mission time series data, in a collaboration between NASA and the European Space Agency (ESA)/Centre National d'Etudes Spatiales (CNES), which is implemented by NExScI. CoRoT is a space telescope operated by the French space agency CNES that takes high-precision photometric data to search for exoplanets and to conduct stellar seismology \citep{auv09}.  The main features of the interface to the CoRoT data are:

\begin{itemize} 
\item Separate and independent interfaces for the exoplanetary and asteroseismology fields;
\item Ability to search by astrophysical or observational parameters, or alternatively by CoRoT ID number;
\item Ability to search across multiple CoRoT runs;
\item Provide direct links to individual CoRoT seismology targets, with the option of accessing any available archive data for the target;
\item Tabular results can be saved for offline analysis;
\item Visualization page for individual CoRoT targets;
\item Download scripts to obtain either all light curves in a given CoRoT run, or to obtain only the light curves that fulfill the specified search criteria.
\end{itemize}

\begin{figure}[h!]
\includegraphics[width=6.5in]{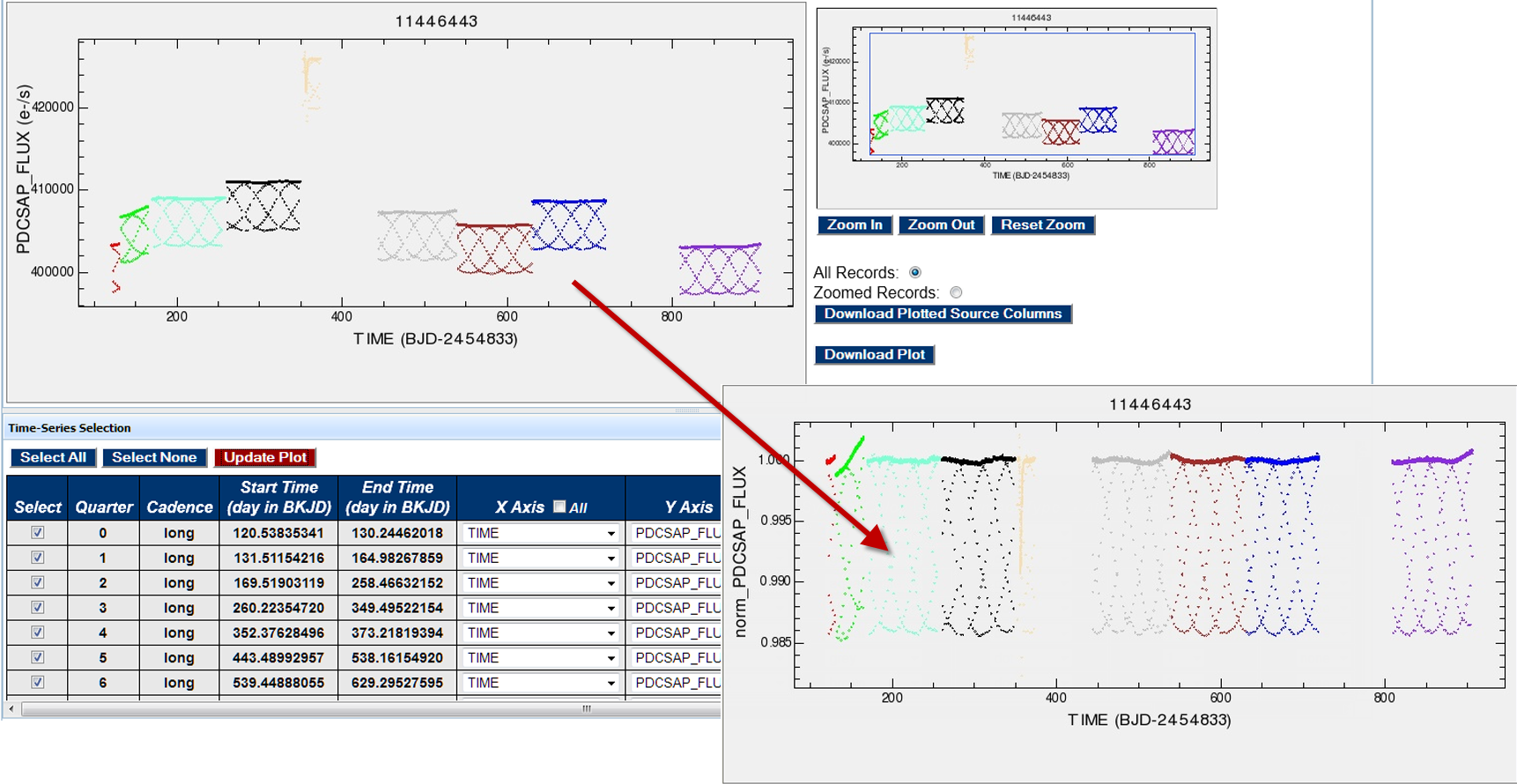}
\caption{\footnotesize Two outputs of the time series viewer demonstrating the normalization capability.
For this example,
quarters 0-9 of data for KIC 11446443 (= KOI 1.01 = Kepler 1 = TrES 2) are plotted.  The top-left plot has selected time on the horizontal axis, 
and PDCSAP\_FLUX (pre-data conditioning simple aperture photometry; a Kepler data product defined as the flux contained in the optimal 
aperture in electrons per second after the PDC module has applied its detrending algorithm to the light curve) on the y-axis.
The small panel in the upper right is the zoom panel; for a full description of the tool capabilities, including detailed instructions on 
normalization, zooming and plot controls, see the user manual at 
http://exoplanetarchive.ipac.caltech.edu/docs/ICETSVhelp.html.
The bottom panel shows the same photometry after it has been normalized by the median value of each individual quarter
(norm\_PDCSAP\_FLUX).  This normalization can be dynamically generated by the user, and is accessed via a tab at the top of the interface (not 
shown).  For this particular light curve shown for KIC 11446443, the peculiar shape of apparently overlapping sinusoids is a byproduct of the 
aliasing of the Kepler ~30 minute cadence with the period of the transiting planet.  It is a Moire pattern resulting from this aliasing that 
disappears when the time series is phased to the correct exoplanet orbital period, as shown in Figure 4.
\label{fig:TSV}
}
\end{figure}

\pagebreak

\subsection{Time Series Properties}

For all Exoplanet Archive survey time series, each data set features a master file and a single file for each light curve.  The master file provides basic properties of the data set as a whole, as well as global parameters about each individual light curve file. Through the archive infrastructure, users may query the master file to search the data set by metadata.  These searchable metadata vary by data set and include unique identifiers, celestial coordinates, static photometry parameters (single-epoch magnitudes), variability filters, observation dates, number of observational epochs, root-mean-square dispersion about the median magnitude, median absolute deviation, existence and frequency of photometric outliers, $\chi^2$ about the median magnitude, etc.  Each light curve file is associated with a unique identifier and features a header summarizing global information about the light curve, as well as the column-delimited photometry data, magnitude, uncertainty. Thus, it is flexible and readable with all computer operating systems and can be translated to other formats such as Virtual Observatory (VO) and binary FITS tables. Figure \ref{fig:TSV} shows an example of data visualization found on the Exoplanet Archive website, complete with light curve characteristics, data set reference, and links to the associated files and download scripts.

\subsection{Ground-based Surveys}

\label{surveys}

The ground-based surveys for transiting exoplanets have been a major contributor to the total yield of known exoplanets. The majority of ground-based transiting exoplanet detections have been provided by the shallow wide-field surveys such as the Transatlantic Exoplanet Survey (TrES) \citep{alo04}, the XO project \citep{mcc05}, the Hungarian Automated Telescope Network (HATNet) \citep{bak04}, SuperWASP \citep{pol06}, and Kilodegree Extremely Little Telescope (KELT) \citep{pep07}.

In addition to the exoplanet discoveries, these surveys have generated an enormous quantity of photometric time series data for stars within the fields surveyed. The Exoplanet Archive stores and serves time series data from the surveys and missions listed in Table \ref{etss}. The interface provides an interactive method for the user to search the data based on positional, photometric, time, and light curve property constraints. Selected time series may be downloaded individually or in bulk using download scripts, as described in \S \ref{bulkdl}. As a note of caution, the Exoplanet Archive does not adjust the time system of the time series as provided by authors of these survey papers.  Combining time series from multiple surveys may result in timing offsets that can produce false transit-timing variation signals.

\begin{deluxetable}{lllll}
  \tablecolumns{5}
  \tablewidth{0pc}
  \tablecaption{\label{surveyholdings} Exoplanet Transit Survey Service Content}
 \tablehead{
    \colhead{Survey Region} &
    \colhead{Objects} &
    \colhead{Time Span} &
    \colhead{Epochs} &
    \colhead{Reference} \\
    \colhead{} &
    \colhead{} &
    \colhead{(days)} &
    \colhead{}
  }
  \startdata
  XO               & 2047   & 486        & $\sim 3500$             &  \citet{mcc05}  \\
  HATNet           & 2656   & 1996       & $\sim1100$--$\sim12000$ & \citet{har11}	\\
  Kepler Field     & 204642 & 1206        & $\sim 500$--$\sim53000$  & \citet{bor10} \\
  CoRoT-Exoplanet  & 125406  & 1411 & 3400--396000            & \citet{auv09} \\
  CoRoT-Seismology & 125     & 1414 & 68000--420000           & \citet{auv09}	\\
  TrES-Lyr1        & 25947  & 75         & $\sim 15000$            &  \citet{alo04} \\
  KELT-Praesepe    & 66637  & 73         & $\sim 3000$             & \citet{pep07}	\\
  NGC 2301         & 3961   & 14         & $\sim 150$              & \citet{how05} \\
  NGC 3201         & 58666  & 700        & $\sim 120$              & \citet{von02} \\
  M 10             & 43930  & 500        & $\sim 50$               & \citet{von02b} \\
  M 12             & 32378  & 500        & $\sim 50$		   & \citet{von02b} \\
  \enddata
\label{etss}
\end{deluxetable}

\subsection{Additional Data}

\label{add:data}

In addition to the project and survey data sets, the Exoplanet Archive includes a wide variety of associated data, such as images \citep[][Two Micron All Sky Survey, 2MASS]{skr06}, spectra, and time series (radial velocity and photometric observations), that can be accessed for each host star.   Other contributed data sets were provided by groups conducting searches for exoplanets.  This includes high-resolution spectra from the N2K Consortium \citep{fis07}, the M2K program \citep{fis12}, and the California Planet search team \citep{wri12}.  These spectra can be accessed for individual planet host stars, or the entire samples provided by the teams can be downloaded in bulk. Infrared spectra as observed by the InfraRed Spectrograph (IRS) on board the Spitzer Space Telescope can be accessed in a similar fashion.

The other categories of additional data in the Exoplanet Archive include high-precision light curves, 
including around 400 radial velocity curves of planet host stars, over 100 photometric planet transit light curves, all photometric light curves from the Hipparcos mission, and about 500 light curves of exoplanet transits as obtained by amateur astronomers from around the world.

All of the ancillary data stored in the Exoplanet Archive is publicly available through electronic versions of publications, or directly contributed by authors.  Information on how to submit data to the Exoplanet Archive is available in the documentation\footnote{\url{http://exoplanetarchive.ipac.caltech.edu/docs/contribute\_data.html}}.

\subsection{Pre-generated Plots}

The current rate of exoplanet discovery is more than 100 new planets per year and increasing, while the KOI list contains more than 2,700 candidates and will continue to increase during the Kepler extended mission. The Exoplanet Archive provides visual aids to allow the user quick access to the latest exoplanet information by creating pre-generated plots of various exoplanet parameters using the currently available data within the archive (Figure \ref{fig:pregen}). Whenever there is an update to the stored information, the plots are re-created to ensure they are as current as possible. These plots provide the community fast access to presentation material that describes the current state of exoplanet research in terms of their number and our understanding of their orbital and physical characteristics. This service is provided for two categories: confirmed exoplanets and Kepler candidates. The plot on the left in Figure 1 shows an example of the confirmed exoplanets that plots the mass (or minimum mass for planets whose signatures have only been detected using the radial velocity method) vs. orbital period. The plot on the right in Figure 1 is a histogram of the discovery rates of exoplanets categorized by discovery method. The histogram shows that 2012 was first year the transit method exceeded the radial velocity method in the number of discoveries.

\section{Tools}

Collecting, vetting and ingesting data is one fundamental aspect of the Exoplanet Archive, and a second is to provide tools for users to display, select, plot and manipulate these data.

\subsection{Interactive Tables}

\label{interactive}

\begin{figure}[h!]
\includegraphics[width=6.5in]{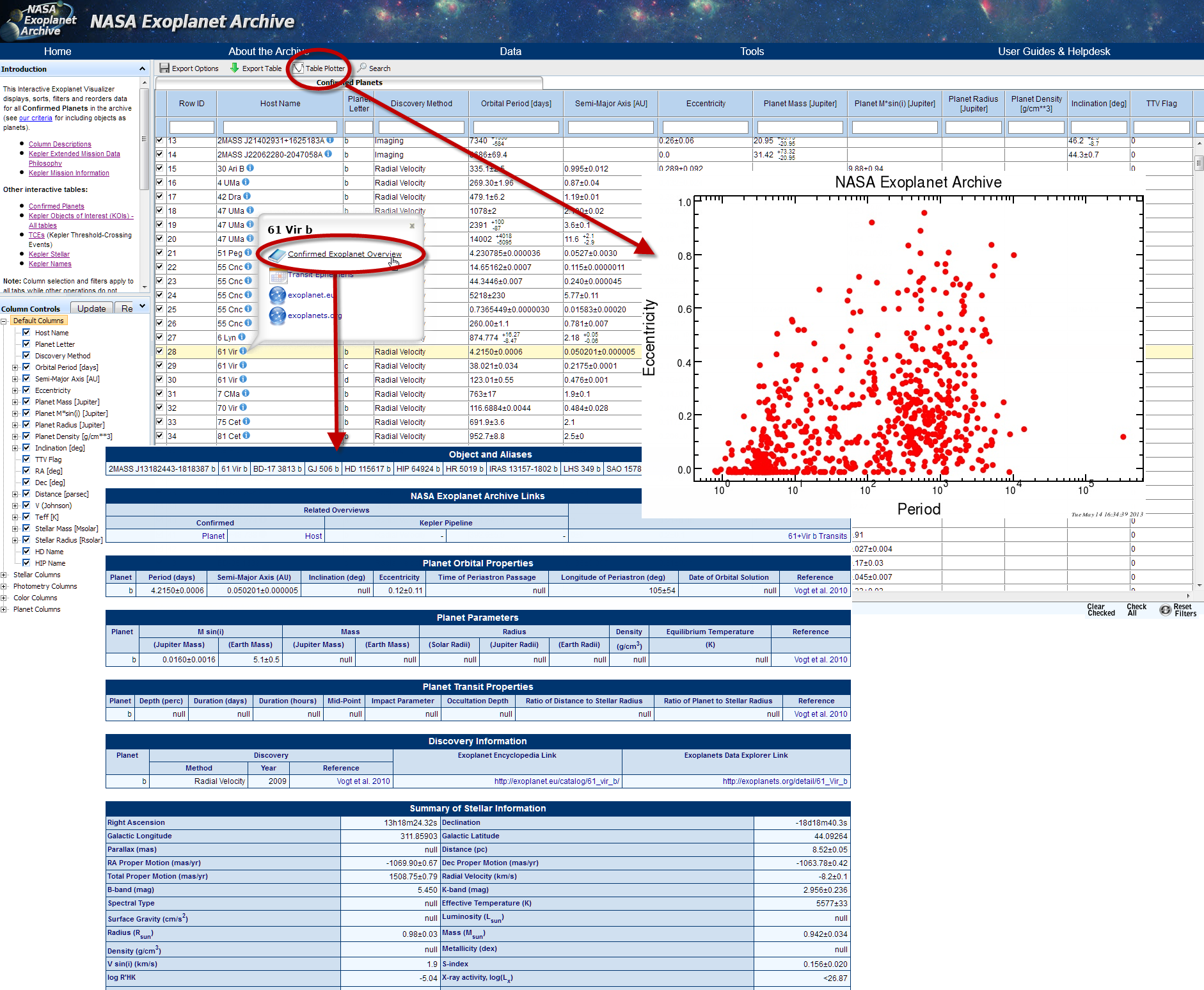}
\caption{\footnotesize A portion of the confirmed planet interactive table showing some of the additional links available from this page, including an overview page for the selected planet, and an interactive table plot for the entire table.   See the user manual at http://exoplanetarchive.ipac.caltech.edu/docs/ICEexohelp.html for more details on table functionality.
\label{fig:table}
}
\end{figure}

As previously discussed, the Exoplanet Archive provides interactive tables for confirmed planets, KOIs, TCEs and target stellar data that allow the user to filter and select data from these sets. This interactive display is also used in the transit service and to display Kepler pipeline products.  In addition to the interactive table features, these tables contain links to other Exoplanet Archive services, such as light curve visualization for the Kepler stars and links to the overview pages described above (Figure \ref{fig:table}).  These links are available by double-clicking on the information icon next to the host star or candidate name.  The flexibility of these tables enables a plethora of user-specified options for selecting only those systems that meet certain criteria. For example, a user with a particular interest in hot Jupiters that transit bright host stars can access these in the Confirmed Planets table by entering $< 5$ in the Orbital Period column, $< 10$ in the V (Johnson) column, and 1 in the Planet Transit Flag column. This will select all transiting planets with orbital periods less than 5 days whose host star is brighter than 10th magnitude. This particular use case could be utilized to select targets for follow-up observations using a specific ground or space-based observatory. The full list of operators available for use in the column filters are described in the table's online documentation\footnote{\url{http://exoplanetarchive.ipac.caltech.edu/docs/ICEexohelp.html\#sortfilter}}. The selection criteria may be far more advanced than this example using the array of available parameters.

The interactive table interface uses an advanced table display and interaction toolkit that is common across most NExScI services.  The client runs on any modern browser, utilizing a collection of JavaScript toolkits (e.g., DHTMLX, jQuery).  Tables are displayed with continuous scrolling, column selection and reorganization, external links, and facilities for defining filters and sorting.  The system uses a server-side database management system for all real table operations, so the client filters translate into full SQL (structured query language) queries.  While this system can handle tables with billions of records, there is a practical limit of $\sim$100,000 records to keep response times reasonable in this interactive mode. The combined service also supports export and plotting capabilities (Figure \ref{fig:table}).

\pagebreak

\subsection{Time Series Viewer}

\label{timeseries}

The Exoplanet Archive time series data visualization service (see Figure \ref{fig:TSV}) allows the user to manipulate and inspect multiple time series (light curves) for a given object.  The primary interface displays a plot of the time series data, allowing the user to select which time series are plotted, per-file or file-independent plotting controls (plotted columns, plot characteristics), and interactive panning and zooming.  Other elements of the service allow the user to compute periodograms and generate phase curve plots while maintaining selection of the time series.  The user can download the raw data files or a JPEG of the displayed plot as well as extract for download an IPAC ASCII table containing the data currently plotted. The service also includes normalization of user-specified columns in the individual time series data files.  The normalized values are available to the other elements of the service and become available for plotting, periodogram calculation and phase curve generation.

As an example, a user might be interested in generating a stitched and normalized light curve for a particular Kepler target to search for a variability signature on time scales longer than the Kepler individual quarter light curve baseline duration of 90 days (for example, KOI 961 = KIC 8561063). This can be accomplished by searching for the Kepler object either through the Kepler light curve search interface, or from the Kepler candidate interactive table for targets that host candidate exoplanets.  Both pathways lead to the lists of individual quarter light curves available for a given Kepler object, where the user can download the data individually or proceed to the time series viewer.

The time series viewer can be used to view individual quarter time series for the object, which will show that both the raw and corrected light curves are often offset from one another in the median flux value. By proceeding to the normalization tab of the time series viewer, the user can easily select all of the individual time series to remove these offsets from within the interface (either via median division or subtraction), and then return to the plot tab of the time series viewer to view the normalized collection of individual quarter time series by selecting the new column containing the normalized flux values. This data product can then be downloaded for offline analysis or sent to the periodogram tool (see \S \ref{periodogram}) for further processing.

\subsection{Periodogram Service}

\label{periodogram}

\begin{figure}[h!]
\includegraphics[width=6.5in]{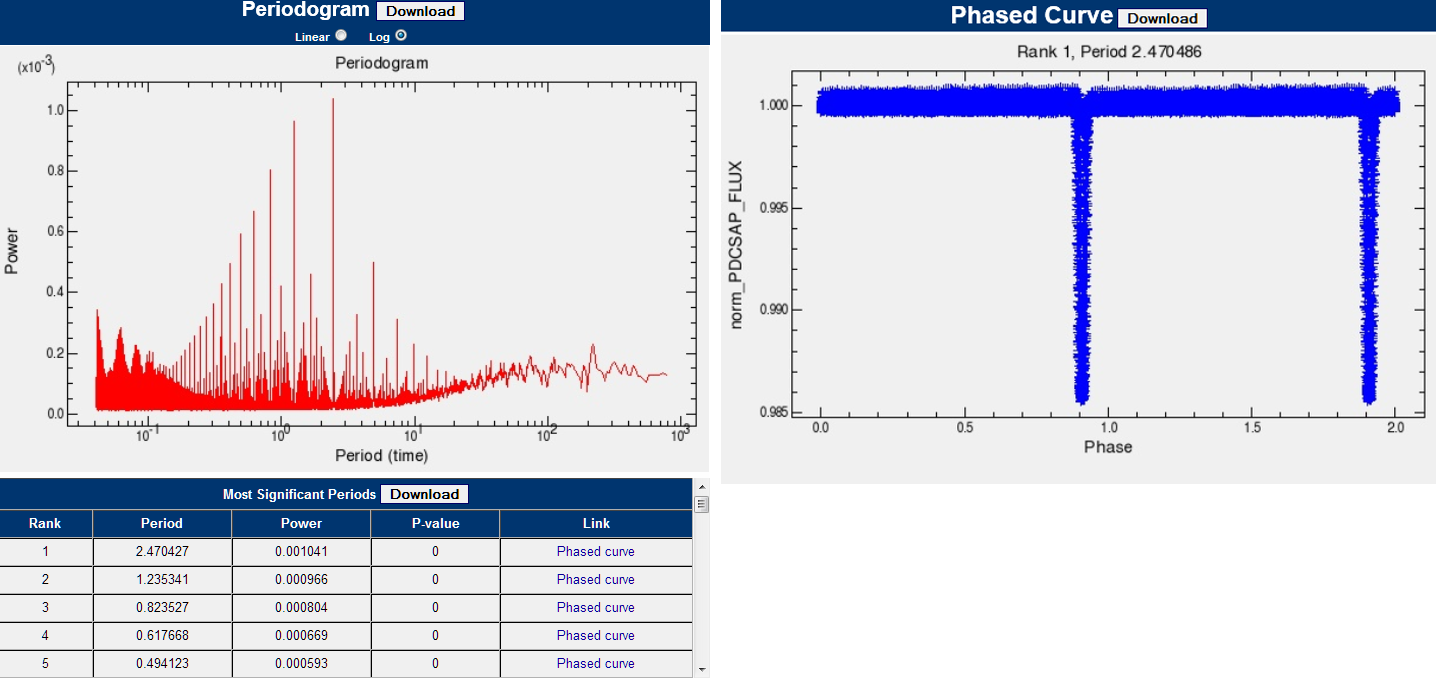}
\caption{\footnotesize 
An example of the periodogram output and corresponding phased light curve. Left:  The BLS periodogram with default parameters 
from the identified period for the object and time series shown in Figure 2 (KIC 11446443 = KOI 1.01 = Kepler 1 = TrES 2). 
The steps to generate this plot are: 
select the Periodogram/Phase Curve tab (not shown in Figure 2) from the time series viewer after computing the normalized time series for quarters 0 to 9, select the norm\_PDCSAP\_flux column for the y-axis, leave the x-axis as Time, select quarters 0 through 9 long cadence photometry from the interface, and click the Compute Periodogram button.  The periodogram type was set to BLS and the default parameters were used.  The horizontal axis is period, and the vertical axis is the power of the period for a given time series (see http://exoplanetarchive.ipac.caltech.edu/applications/Periodogram/docs/Algorithms.html for more information on how the power is computed for a given algorithm.) For this particular time series for KIC 11446443, there are many of peaks in the periodogram.  Many of these are aliases of the fundamental period, which in this case is the highest-ranked period as shown below the plot.  The table in the bottom left displays the highest-ranked periods, their power values and associated P-values (false-alarm probabilities).  Again, the p-values should be interpreted with caution as described on our Algorithms page.  The Link column contains a link to the phased time series, which is shown in the right panel at a period of 2.470527 days.
\label{fig:period}
}
\end{figure}

The periodogram service in the Exoplanet Archive returns periodograms of time series data either from data in the archive or data uploaded by the user. Periodograms are computed to extract periodic signals from time series data. Calculating a periodogram is similar to computing a Fourier transform, in which the input data are transformed from the time domain to the frequency domain. The power at a candidate frequency indicates the strength of a repeating signal in the data at that frequency. The periodogram code returns two important results: the periodogram itself (the spectral power as a function of frequency) and a table of the peaks in the periodogram, with the associated probabilities that these peaks arise by chance (Figure \ref{fig:period}). A variety of astrophysical phenomena can produce periodic time series curves, including stellar variability, eclipsing binaries and transiting planets. Time series curves of astrophysical objects are inherently noisy measurements;  photon noise, atmospheric conditions, instrumental artifacts, systematic trends, and other factors can introduce random variation into the photometric magnitude of the observations. The intervals at which the measurements are taken are also generally uneven and periodograms are particularly useful to extract intrinsic periodic signals in this case.

The service supports three algorithms: Lomb-Scargle \citep{sca82,hor86,zec09}, box-fitting least squares \citep[BLS;][]{kov02} and Plavchan \citep{pla08,ste78}.  
By default, the service uses the Lomb-Scargle algorithm. Users may select the BLS and Plavchan algorithms and change parameters from the results page.  The Lomb-Scargle algorithm is widely used by the community and is most helpful for identifying sinusoidal-like periodic variations for irregularly sampled data, as is common in astrophysical data sets when a direct Fourier transform cannot be computed. The BLS algorithm is optimized for the detection of periodic transit- or eclipse-shaped events (i.e., repeating ``top-hats'' or  ``boxes''), and has been used successfully in the detection of many transiting exoplanets listed in the Exoplanet Archive.  As a note of caution, for eclipsing binaries and especially equal mass eclipsing binaries, the BLS algorithm will often report the half-period of the binary as the most significant period and the results should be carefully interpreted.  Finally, the Plavchan algorithm is a bin-less variation of the phase-dispersion minimization approach to finding periodic variability in a time series.  It can identify both sinusoidal-like and transit-like periodic variability, but also everything in between, including sawtooth shapes and RR Lyrae-type light curves, at the expense of extra computational time.  We refer the reader to the documentation available on these algorithms at the archive\footnote{\url{http://exoplanetarchive.ipac.caltech.edu/applications/Periodogram/docs/Algorithms.html}}.

The service includes a number of adjustable parameters such as the period range and algorithm-specific parameter. The periodogram service is integrated into the time series viewer for the Kepler light curves and is available for all light curves in the Exoplanet Archive, but will also accept user-uploaded data files.  The Exoplanet Archive also provides tools to convert data formats into one compatible with the periodogram service. Visualization is provided for user-submitted time series data, giving the user an opportunity to view the light curve prior to sending it to the periodogram service. The periodogram can be a very powerful tool for automating the identification of statistically significant periods of variation in time series. However, one must exercise caution in interpreting the results table of the most significant periods.  The calculated statistical significance (p-value, the probability of a false-positive period) of ranked periods may not be reliable. Several factors may invalidate the assumptions applied in estimating the statistical significance, including changes to the input parameters such as the period minimum and maximum search, as well as the number of periods search and how those periods are calculated (for example, constant period steps, or constant steps in frequency, under-sampling vs. over-sampling, etc.). 

The periodogram service calculations utilize a 128-node cluster at NExScI. The processing is partitioned into a back-end, which does the transform for a set of frequencies, and a front-end, which handles the logic of splitting the processing by frequency ranges and combining the results into the final periodogram. In the ideal case, the processing would speed up by a factor equal to the number of processors. In practice, there are various sources of overhead, such as managing the distributed processing and collecting and combining the results into a single file, resulting in a small decrease in efficiency.

An example use for this tool is a search for the periodic variability signature due to an eclipsing binary.  Once loaded with a time series, the service attempts an initial guess for the optimal period searching parameter -- minimum and maximum periods, the frequency step, etc. -- based on the time baseline and number of data points in the time series. By default, the periodogram service runs a period search using the Lomb-Scargle algorithm, but in this example, the Lomb-Scargle algorithm is not optimal.  Either the BLS or Plavchan algorithms may be more appropriate.  The periodogram tool can be configured to select data from multiple columns, and the period search parameters and algorithm specific settings can also be changed. The user can confirm the accuracy of the identified periods of variability in the results returned by the periodogram service by investigating the associated p-values, and by visually inspecting the phased time series.  Occasionally, the periodogram will identify the half-period for the eclipsing binary. If this occurs, the user can adjust the period minimum and maximum from the defaults to constrain the period search around twice the identified period from the initial search, perhaps with smaller frequency steps.  Then the resulting periodogram can be downloaded to estimate the period uncertainty from the width of the peak in the periodogram, as well as used to perform a different estimate of the false alarm probability from the one used in the processing.

\subsection{Transit Ephemeris Service}

The transit service in the Exoplanet Archive is a tool for predicting future transits and is based on a service developed by Greg Laughlin \footnote{\url{http://transitsearch.org}}. The service queries two tables (updated nightly) that contain information on the next transit for all known transiting planets and KOIs. The service can be run for an individual, confirmed planet to predict several forward transits, using either orbital or transit parameters gathered from the Exoplanet Archive or provided by the user, and can also calculate all transits visible for a specified time range and location (viewable transit service). To facilitate the planning of follow-up observations, the viewable transit service is pre-loaded with the latitude and longitude of various ground-based observatories and also supports custom coordinates entered by users. This tool also provides the capability to identify transit events that fall within the visibility windows for the Spitzer Space Telescope.  The transit tools can be accessed either from the Exoplanet Archive front page (Transit and Ephemeris Predictor and Viewable Transit Service) or from the links within the interactive tables for confirmed planets and KOIs.  Any changes in parameters for either the confirmed planets or the KOIs are immediately available in the calculated transits and planned updates include display of the quadrature points and plots of the viewable transits. For confirmed planets that are not known transiters or do not have enough orbital information to predict transits, the tool will identify the data parameters the user must supply in order to predict future transits. Currently, the transit calculations do not take into account any known Transit Timing Variations (TTVs) when predicting transit times.

\subsection{Bulk Data Download}

\label{bulkdl}

The Exoplanet Archive provides wget scripts\footnote{\url{http://irsa.ipac.caltech.edu/docs/batch\_download\_help.html}} that enable users to download a variety of spectra, time series and other data for stars known to host exoplanets. Light curves and spectra are served from the Hipparcos mission, the N2K Consortium, the M2K program, the California Planet Search team, and amateur astronomers around the globe. These include radial velocity and photometric data related to planets detected by a variety of techniques. Also available are high-precision data from photometric surveys for transiting exoplanets from various projects and missions, including CoRoT, Kepler, and the ground-based data listed in Table 1.  The Kepler data validation reports and summaries are also available via wget scripts.  For a complete listing of all of the data available via bulk download, see \url{http://exoplanetarchive.ipac.caltech.edu/bulk\_data\_download/}.

By removing rows, downloaded wget scripts can easily be customized by users to fetch only specific data of interest.  This is particularly useful for users who would like to conduct their own statistical studies and/or object searches in these data, such as an independent search for planetary transit signatures in the photometric survey data.   The wget scripts work on
Windows, Mac and Linux operating systems.

\subsection{Application Program Interface}

The Exoplanet Archive provides an easy-to-use application program interface (API) that allows users to access and download
parameter values as contained in the interactive tables, including the confirmed exoplanet list and the Kepler KOI, stellar and TCE data.   Simple URLs are constructed and submitted from a browser to retrieve data.  Submitted URLs specify the table to be accessed, the columns to be returned, the format of the returned data, and any specified "where" clauses.  Once users have narrowed down the data of interest  to a simple query, they can submit it regularly to get the most up-to-date data. This means the user can create a custom API query to check the database periodically and retrieve the data specific to the user's science goals, such as extracting the stellar information for exoplanet host stars for creating target lists.

Documentation and examples for API usage can be found in the online user documentation (\url{http://exoplanetarchive.ipac.caltech.edu/docs/program\_interfaces.html}).

\subsection{Documentation and User Help}

\label{dox}

The Exoplanet Archive includes extensive documentation, both to explain the data contents and to detail how to use the tools. Each interactive table has a listing of all columns with units and descriptions.  Each tool has a user guide with examples, screen shots and known issues. Questions and suggestions can be sent to the Exoplanet Archive team via a web form on the website\footnote {\url{http://exoplanetarchive.ipac.caltech.edu/cgi-bin/Helpdesk/nph-genTicketForm}}. Additionally, users can contact us on our Facebook page\footnote{\url{https://www.facebook.com/NASAExoplanetArchive}}, Google+\footnote{\url{https://plus.google.com/116475294526790703237}}, watch tutorials on YouTube\footnote{\url{http://www.youtube.com/user/NASAExoplanetArchive}} or join our mailing list\footnote{\url{https://lists.ipac.caltech.edu/mailman/listinfo/exoplanet-announce}}.

\section{Summary and Future Work}

The Exoplanet Archive is designed and operated to facilitate exoplanet research by serving as a repository for planetary and stellar physical and orbital properties, and by providing tools to work with these data along with light curves from Kepler, CoRoT and ground-based surveys. As one of the main goals is to compile exoplanet and host star data from the refereed literature, the contents of the Exoplanet Archive are updated almost every week, both with newly discovered planets and with additional data for objects already in the archive. The Exoplanet Archive also hosts Kepler pipeline data, including planet candidate lists that are updated as often as weekly, pipeline-identified threshold-crossing events (TCEs), data validation documentation and target stellar data. We strongly encourage users to work with us to enhance, update and make corrections to the content of the Exoplanet Archive by contributing data or sending corrections and suggestions.

The Exoplanet Archive team is actively working to increase the interoperability of tools, such as using the same time series viewer for all light curves.  Other areas of development include adding the ability for users to work with time series data with statistics, binning and whitening calculations, and storage of user queries and preferences.

\section{Acknowledgements}

The Exoplanet Archive is funded through NASA's Exoplanet Exploration Program, administered by the Jet Propulsion Laboratory, California Institute of Technology. This publication makes use of data products from the Two Micron All Sky Survey, which is a joint project of the University of Massachusetts and the Infrared Processing and Analysis Center/California Institute of Technology, funded by the National Aeronautics and Space Administration and the National Science Foundation. This research has made use of the NASA's IPAC Infrared Science Archive, which is operated by the Jet Propulsion Laboratory, California Institute of Technology, under contract with the National Aeronautics and Space Administration. The Kepler light curves were obtained from the Mikulski Archive for Space Telescopes (MAST). STScI is operated by the Association of Universities for Research in Astronomy, Inc., under NASA contract NAS5-26555. Support for MAST for non-HST data is provided by the NASA Office of Space Science via grant NNX09AF08G and by other grants and contracts. This research has made use of the SIMBAD database, operated at CDS, Strasbourg, France.

The Exoplanet Archive gratefully acknowledges the contributions of many in the community who have provided data for the archive and who work to maintain resources for the exoplanet community.   We thank the anonymous referee who provided many helpful comments and suggestions.


\begin{thebibliography}{}

\bibitem[Alonso et al.(2004)]{alo04} Alonso, R., Brown, 
T.~M., Torres, G., et al.\ 2004, \apjl, 613, L153 

\bibitem[Auvergne et 
al.(2009)]{auv09} Auvergne, M., Bodin, P., Boisnard, L., et al.\ 2009, \aap, 506, 411 


\bibitem[Bakos et al.(2004)]{bak04} Bakos, G., Noyes, R.~W., 
Kov{\'a}cs, G., et al.\ 2004, \pasp, 116, 266 

\bibitem[Batalha et al.(2013)]{bat13} Batalha, N.~M., Rowe, 
J.~F., Bryson, S.~T., et al.\ 2013, \apjs, 204, 24 

\bibitem[Bean et al.(2010)]{bea10} Bean, J.~L., Seifahrt, A., 
Hartman, H., et al.\ 2010, \apjl, 711, L19 

\bibitem[Borucki et al.(2011)]{bor11b} Borucki, W.~J., Koch, 
D.~G., Basri, G., et al.\ 2011, \apj, 736, 19 

\bibitem[Borucki et al.(2011)]{bor11a} Borucki, W.~J., Koch, 
D.~G., Basri, G., et al.\ 2011, \apj, 728, 117 

\bibitem[Borucki et al.(2010)]{bor10} Borucki, W.~J., Koch, 
D., Basri, G., et al.\ 2010, Science, 327, 977 



\bibitem[Cannon 
\& Pickering(1918)]{can18} Cannon, A.~J., \& Pickering, E.~C.\ 1918, Annals of Harvard College Observatory, 91, 1 


\bibitem[Debosscher et 
al.(2011)]{deb11} Debosscher, J., Blomme, J., Aerts, C., \& De Ridder, J.\ 2011, \aap, 529, A89 

\bibitem[Fischer et al.(2007)]{fis07} Fischer, D.~A., Vogt, 
S.~S., Marcy, G.~W., et al.\ 2007, \apj, 669, 1336 

\bibitem[Fischer et al.(2012)]{fis12} Fischer, D.~A., Gaidos, 
E., Howard, A.~W., et al.\ 2012, \apj, 745, 21 


\bibitem[Hartman et al.(2011)]{har11} Hartman, J.~D., Bakos, 
G.~{\'A}., Noyes, R.~W., et al.\ 2011, \aj, 141, 166 


\bibitem[Horne 
\& Baliunas(1986)]{hor86} Horne, J.~H., \& Baliunas, S.~L.\ 1986, \apj, 302, 757 

\bibitem[Howell et al.(2005)]{how05} Howell, S.~B., 
VanOutryve, C., Tonry, J.~L., Everett, M.~E., 
\& Schneider, R.\ 2005, \pasp, 117, 1187 

\bibitem[Jenkins et al.(2010)]{jen10} Jenkins, J.~M., 
Caldwell, D.~A., Chandrasekaran, H., et al.\ 2010, \apjl, 713, L87 


\bibitem[Kov{\'a}cs et al.(2002)]{kov02} Kov{\'a}cs, G., Zucker, S., \& Mazeh, T.\ 2002, \aap, 391, 369 

\bibitem[Mayor 
\& Queloz(1995)]{may95} Mayor, M., \& Queloz, D.\ 1995, \nat, 378, 355 

\bibitem[McCullough et al.(2005)]{mcc05} McCullough, P.~R., 
Stys, J.~E., Valenti, J.~A., et al.\ 2005, \pasp, 117, 783 

\bibitem[Nymeyer et al.(2011)]{nym11} Nymeyer, S., 
Harrington, J., Hardy, R.~A., et al.\ 2011, \apj, 742, 35 

\bibitem[Pepper et al.(2007)]{pep07} Pepper, J., Pogge, 
R.~W., DePoy, D.~L., et al.\ 2007, \pasp, 119, 923 

\bibitem[Plavchan et al.(2008)]{pla08} Plavchan, P., Jura, 
M., Kirkpatrick, J.~D., Cutri, R.~M., 
\& Gallagher, S.~C.\ 2008, \apjs, 175, 191 

\bibitem[Poddan{\'y} et al.(2010)]{pod10} Poddan{\'y}, S., 
Br{\'a}t, L., \& Pejcha, O.\ 2010, \na, 15, 297 

\bibitem[Pollacco et al.(2006)]{pol06} Pollacco, D.~L., 
Skillen, I., Collier Cameron, A., et al.\ 2006, \pasp, 118, 1407 

\bibitem[Pravdo 
\& Shaklan(2009)]{pra09} Pravdo, S.~H., \& Shaklan, S.~B.\ 2009, \apj, 700, 623 


\bibitem[Prsa et al.(2011)]{prs11} Pr{\v s}a, A., 
Batalha, N., Slawson, R.~W., et al.\ 2011, \aj, 141, 83 


\bibitem[Ramirez et al.(2009)]{ram09} Ramirez, S., Ali, B., 
Baker, R., et al.\ 2009, IAU Symposium, 253, 474 

\bibitem[Scargle(1982)]{sca82} Scargle, J.~D.\ 1982, \apj, 
263, 835 

\bibitem[Schneider et 
al.(2011)]{sch11} Schneider, J., Dedieu, C., Le Sidaner, P., Savalle, R., \& Zolotukhin, I.\ 2011, \aap, 532, A79 

\bibitem[Skrutskie et al.(2006)]{skr06} Skrutskie, M.~F., 
Cutri, R.~M., Stiening, R., et al.\ 2006, \aj, 131, 1163 

\bibitem[Stellingwerf(1978)]{ste78} Stellingwerf, R.~F.\ 
1978, \apj, 224, 953 

\bibitem[Swift et al.(2012)]{swi12} Swift, D.~C., Eggert, 
J.~H., Hicks, D.~G., et al.\ 2012, \apj, 744, 59 

\bibitem[von Braun 
\& Mateo(2002)]{von02} von Braun, K., \& Mateo, M.\ 2002, \aj, 123, 279 

\bibitem[von Braun et al.(2002)]{von02b} von Braun, K., Mateo, 
M., Chiboucas, K., Athey, A., \& Hurley-Keller, D.\ 2002, \aj, 124, 2067 

\bibitem[von Braun et al.(2009)]{von09} von Braun, K., 
Abajian, M., Ali, B., et al.\ 2009, IAU Symposium, 253, 478 

\bibitem[Wright et al.(2012)]{wri12} Wright, J.~T., Marcy, 
G.~W., Howard, A.~W., et al.\ 2012, \apj, 753, 160 

\bibitem[Wright et al.(2011)]{wri11} Wright, J.~T., Fakhouri, 
O., Marcy, G.~W., et al.\ 2011, \pasp, 123, 412 

\bibitem[Zechmeister \& Kurster(2009)]{zec09} Zechmeister, M.,  Kurster, M.\ 2009, \aap, 496, 577 


\end{thebibliography}
\end{document}